\documentclass[12pt, conference, compsocconf, onecolumn]{IEEEtran}
\ifCLASSINFOpdf
  \usepackage[pdftex]{graphicx}
\else
  \usepackage[dvips]{graphicx}
\fi

\usepackage{pgfplots}
\usepackage{booktabs}
\usepackage{hyperref}
\usepackage[cmex10]{amsmath}

\pgfplotsset{width=10cm,compat=1.9, tick align=outside}
\pgfplotscreateplotcyclelist{my exotic colorlist}{%
  teal,every mark/.append style={fill=teal!40!black},mark=otimes*\\%
  orange,every mark/.append style={fill=orange!80!black},mark=square*\\%
  cyan!60!black,every mark/.append style={fill=cyan!80!black},mark=triangle*\\%
  red!70!white,mark=star\\%
  lime!80!black,every mark/.append style={fill=lime},mark=diamond\\%
  red,every mark/.append style={solid,fill=red!80!black},mark=*\\%
  yellow!60!black,densely dashed,every mark/.append style={solid,fill=yellow!80!black},mark=square*\\%
  black,every mark/.append style={solid,fill=gray},mark=otimes*\\%
  blue,densely dashed,mark=star,every mark/.append style=solid\\%
  red,densely dashed,every mark/.append style={solid,fill=red!80!black},mark=diamond*\\%
}

\title{Robust 3D U-Net Segmentation of Macular Holes}

\author{\IEEEauthorblockN{Jonathan Frawley\IEEEauthorrefmark{1}\IEEEauthorrefmark{2}, Chris G.~Willcocks\IEEEauthorrefmark{1}, Maged Habib\IEEEauthorrefmark{3}, \\ Caspar Geenen\IEEEauthorrefmark{3}, David H.~Steel\IEEEauthorrefmark{3}\IEEEauthorrefmark{4} and Boguslaw Obara\IEEEauthorrefmark{1}\IEEEauthorrefmark{2}}
\IEEEauthorblockA{\IEEEauthorrefmark{1}\textit{Department of Computer Science}, \textit{Durham University}, Durham, UK}
\IEEEauthorblockA{\IEEEauthorrefmark{2}\textit{Intogral Limited}, Durham, UK}
\IEEEauthorblockA{\IEEEauthorrefmark{3}\textit{Sunderland Eye Infirmary}, Sunderland, UK}
\IEEEauthorblockA{\IEEEauthorrefmark{4}\textit{Newcastle University}, Newcastle Upon Tyne, UK}
}

\begin{document}

\maketitle

\begin{abstract}
Macular holes are a common eye condition which result in visual impairment. We look at the application of deep convolutional neural networks to the problem of macular hole segmentation.
We use the 3D U-Net architecture as a basis and experiment with a number of design variants.
Manually annotating and measuring macular holes is time consuming and error prone, taking dozens of minutes to annotate a single 3D scan.
Previous automated approaches to macular hole segmentation take minutes to segment a single 3D scan.
We found that deep learning models generate significantly more accurate segmentations than previous automated approaches (Jaccard index boost of $0.08-0.09$) and expert agreement (Jaccard index boost of $0.13-0.20$) in less than one second.
We also demonstrate that an approach of architectural simplification, by greatly simplifying the network capacity and depth, results in a model which is competitive with state-of-the-art models such as residual 3D U-Nets.
\end{abstract}


\section{Introduction}
\label{sec:intro}
Idiopathic full thickness macular holes (iFTMH) are a common, and visually disabling condition, being bilateral in 10\% of affected individuals.
They occur at a prevalence of approximately 1 in 200 of the over 60-year-old population with an incidence of approximately 4000 per annum in the United Kingdom (UK)\cite{ali2017incidence}\cite{mccannel2009population}.
If left untreated they result in visual acuity below the definition of blindness and typically greater than 1.0 logMAR (logarithm of the minimum angle of resolution), where 0.1 logMAR is classed as normal. 

3D high-resolution images of the retina can be created using optical coherence tomography (OCT)~\cite{hee1995optical}.
It is now the standard tool for diagnosing macular holes \cite{goldberg2014optical}.
Compared to previous imaging methods, OCT can more easily assist a clinician in differentiating a full-thickness macular hole from mimicking pathology which is important in defining appropriate treatment \cite{hee1995optical}.
An OCT scan of a macular hole is a 3D volume.
Clinicians, however, typically view OCT images as a series of 2D images, choose the central slice with maximum dimensions and perform measurements which are predictors of anatomical and visual success such as base diameter, macular hole inner opening and minimum linear diameter \cite{madi2016optimal}\cite{chen2020macular}\cite{murphy2020}\cite{steel2020factors}.
This approach is limited as it assumes that the macular hole base is circular, and would give incorrect results when it is elliptical \cite{Nasrulloh2018}, which is typically the case~\cite{chen2020macular}.
With the advent of automated 3D approaches, it is possible to begin to look at measurements in 3D and how they might be predictors of anatomical and visual success.

Neural networks are an interconnected group of artificial neurons, which can be reconfigured to solve a problem based on data.
Convolutional neural networks (CNN) are a type of neural network inspired by how the brain processes visual information~\cite{Lindsay_2020}.
CNNs have been very successful in computer vision problems, such as automating the segmentation of medical images.
For a CNN to learn to segment images in a supervised manner, it needs to have access to images with associated ground truth (GT) information which highlight the areas of the image for the task at hand.
This is often done manually which is time consuming and requires expert knowledge.
The U-Net CNN architecture~\cite{ronneberger2015u} is a highly-utilized CNN architecture for biomedical image segmentation.
It has had success in segmentation to help diagnose other eye diseases such as macular edema, even when dataset sizes are limited~\cite{frawleybibe2020}.
We sought to examine the application of variants of the U-Net architecture to the problem of macular hole segmentation.
Our proposed model is a smaller version of the model from the original 3D U-Net paper~\cite{cciccek20163d}.
We also implemented and evaluated the proposed model with residual blocks added, similar to those described by He et al.~\cite{he2016identity}.
In addition, we implemented a much more complex residual model, DeepMind's OCT segmentation model~\cite{de2018clinically}, and ran the same tests with it.

Our contribution can be summarised as developing an automated approach to macular hole segmentation based on deep learning which yields significantly improved results compared to prior methods.
We present a comparison of the above-mentioned models against the current state-of-the-art automated approach~\cite{Nasrulloh2018} along with a comparison against expert agreement.
The state-of-the-art method is a level set approach which does not use deep learning.
We show that simple low-capacity 3D U-Nets are capable of outperforming the state-of-the-art automated approach and that increasing the complexity of the architecture does not improve performance.

\begin{figure*}[!t]
  \centering
  \includegraphics[width=\linewidth]{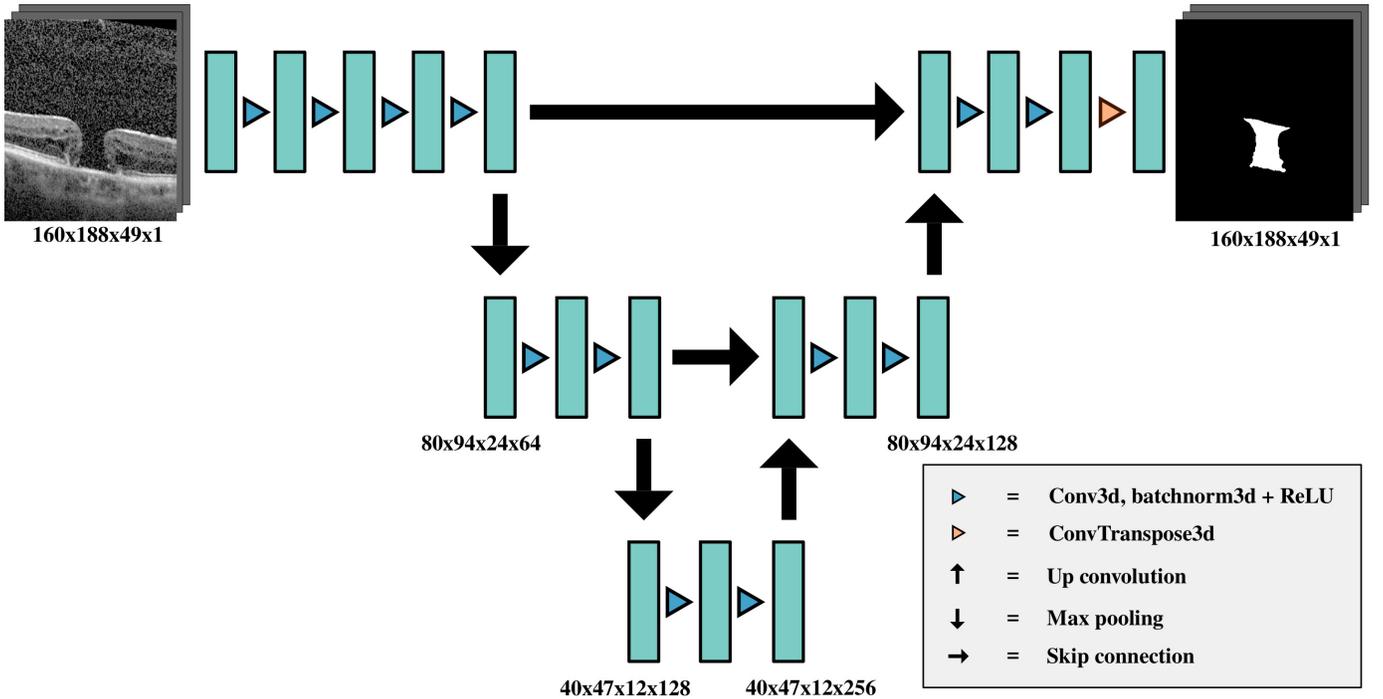}
  \caption{Small 3D U-Net ($M_{1}$). The proposed model is a cut-down version of 3D U-Net~\cite{cciccek20163d}. It has fewer levels and a carefully optimized capacity for our datasets.}
  \label{fig:deeplearningmodel}
\end{figure*}

\section{Method}
Image segmentation involves the labelling of objects of interest in an image.
For a 3D image, this is done by assigning voxels with shared characteristics to corresponding class labels.
We wish to assign areas of the macular hole volume in an OCT image to white voxels and all other regions to black voxels.

We use binary cross-entropy as our loss function, which tells us how close our predicted macular hole regions are to those in the ground truth:
\begin{equation}
\begin{split}
    \mathcal{L}_{\text{BCE}} & = - \dfrac{1}{N} \sum_{i=1}^{N} p_i \log q_i + (1-p_i) \log (1-q_i),
\end{split}
\end{equation}
$N$ being the batch size, $p_i$ being the ground truth and $q_i$ being the output of our model.
For images with multiple annotations in our training set, we trust them with equal integrity and the target probabilities are averaged.
The validation set has no samples with multiple annotations.
Originally we had three annotations for each OCT image in the unseen test set.
However, due to inconsistencies between authors, we combined all ground truths into a single ground truth per image.
To do this, we used a voting system, where if $\frac{2}{3}$ of the authors had annotated a voxel, that voxel was annotated in the resultant ground truth.
Experiments using random scaling and elastic deformation augmentation yielded no improvement in performance and we did not use any data augmentation in our final experiments.

U-Net takes as input a 2D image and outputs a set of probabilities.
Each entry in the output is the probability of each part of the image being a part of the segmented region.
It is a U-shaped CNN architecture, consisting of a contracting path and an expansive path.
The contracting path consists of 2D convolutions, ReLU activations and 2D max pooling at each level.
The expansive path's levels use \emph{skip connections} to their contracting path equivalent, along with up convolutions and ReLU activations.
Skip connections allow for high-resolution information to be captured by the model while the contracting/expansive paths capture the abstract shape of the segmentation.
The 3D U-Net architecture~\cite{cciccek20163d} is a version of U-Net designed for use with 3D images which uses 3D convolutions, up convolutions and max pooling layers.
This allows for improved segmentation of 3D images as the context from multiple slices are used to decide whether an individual voxel is an object or not.

A number of models based on the 3D U-Net architecture were compared:
\begin{enumerate}
  \setlength\itemsep{0em}
  \addtolength{\itemindent}{1em}
  \item [$M_{1}$:] Small 3D U-Net (Proposal) [5,216,353 parameters]
  \item [$M_{2}$:] Small residual 3D U-Net (Residual) [13,928,833 parameters]
  \item [$M_{3}$:] Residual 3D U-Net for 2D slices (DeepMind)~\cite{de2018clinically} [470,333,089 parameters]
\end{enumerate}

A diagram of model $M_{1}$ is shown in \figurename~\ref{fig:deeplearningmodel}.
Our experiments showed that using three levels for this model resulted in the best performance.
A scaled-down input image of $160 \times 188 \times 49$ yielded the best results for models $M_{1}$ and $M_{2}$.
The output is of the same dimensions as the input.
$M_{2}$ is similar to $M_{1}$ except that residual blocks have been added to each level.
$M_{3}$ is a residual 3D U-Net architecture which takes nine slices of the OCT image as input and outputs a 2D probability map as output, representing the segmentation of a single slice of the OCT image.
For $M_{3}$, the slice which we want to segment, along with 4 slices on either side is input to the model, which is a $321 \times 376 \times 9$ image.
For slices near the boundaries, we use mirroring to handle slices which are outside of the image.
It outputs a set of $321 \times 376$ probabilities, corresponding to one slice of the 3D OCT.
$M_{3}$, therefore, requires 49 iterations to segment a whole 3D OCT image in our dataset.
Model $M_{3}$ has the most parameters of the models tested, with $M_{1}$ having the fewest parameters.

Jaccard index was used as the primary metric for measuring the performance of each method.
This is one of the standard measures of the performance of image segmentation methods, especially in medical image segmentation~\cite{taha2015metrics}.
Dice similarity coefficient (DSC) is also commonly used and is closely related to the Jaccard index, with one being computable from the other.
For completeness and ease of comparison, we also provide DSC for our proposed model in Section~\ref{sec:quant}.

\section{Materials}
OCT images were exported from a Heidelberg SPECTRALIS (Heidelberg, Germany) machine.
A high density central horizontal scanning protocol with 29-30 micron line spacing was used in the central 15 by 5 degrees.
The individual OCT line scans had spacing which varied slightly between images but it was typically 5.47 microns per pixel on the $x$-axis and 3.87 microns per pixel on the $y$-axis.
All scans used a 16 automatic real-time setting enabling multisampling and noise reduction over 16 images. 
All images were cropped to the same size and unnecessary information such as the fundus image were removed.
Annotations were created by a mixture of clinicians and image experts using a 3D image annotation tool.
Pixels on each slice of the OCT scan which represented macular hole were highlighted.
There were 85 (image, annotation) pairs in the training dataset, 56 after combining annotations from multiple authors.
There were 22 pairs in the validation dataset and 9 in the unseen test set.
All images and ground truths at full size had dimensions $321 \times 376 \times 49$.

\section{Implementation}
\label{sec:impl}
Our experiments were all conducted using the PyTorch~\cite{pytorch} deep learning framework on NVIDIA Turing GPUs with $24\text{GB}$ of memory.
We trained each model for $500$ epochs where each epoch ran over 10 3D images.
This means that the models which output a 3D segmentation ($M_{1}$ and $M_{2}$) had 10 iterations per epoch, and the slice-based model ($M_{3}$) had 490 iterations per epoch.
As source code was not released for DeepMind's model, $M_{3}$ is implemented as closely as possible from the description provided in the original paper and slightly adapted to fit the binary classification problem.

In order to evaluate models $M_{1}$ and $M_{2}$, we scaled up the output probability map to its original size using trilinear interpolation and thresholded it at $0.5$ to generate a binary segmentation.
For model $M_{3}$, we individually ran over all 49 slices of an image and recombine the 49 2D probability maps into a single 3D probability map.
We then thresholded this combined map at $0.5$ to generate a 3D binary segmentation.
The Adam optimization algorithm~\cite{kingma2014adam} was used to optimize parameters of the models, with hyperparameters being found by experimentation.
The \emph{BCEWithLogitsLoss} function in PyTorch was used for loss calculation, which combines a sigmoid activation and binary cross entropy loss into one function.
A similar number of experiments were conducted for each model.
For model $M_{1}$, a learning rate of $1\mathrm{e}{-4}$ and weight decay of $1\mathrm{e}{-6}$ was used.
For model $M_{2}$, a learning rate of $1\mathrm{e}{-4}$ and weight decay of $1\mathrm{e}{-5}$ was used.
For model $M_{3}$, a learning rate of $7.5\mathrm{e}{-5}$ was used and weight decay was disabled.
The 3D OCT images were normalized to the $\left[0,1\right]$ range prior to scaling or slicing.

Each model was trained and evaluated separately three times to assess the consistency of our results.
Due to the fact that we only had a small number of images with multiple authors, we decided to keep the training, validation and unseen test set static for all tests rather than using k-fold cross-validation.
We reserved all images which had three annotations for the unseen test set, in order for us to be able to compare our results with expert agreement, which was a key goal of the research.

\begin{figure*}[t]
  \setlength\tabcolsep{.001\linewidth} 
  \centering
  \begin{tabular}{cccccc}
    \includegraphics[trim=50 110 50 100, clip, width=.166\linewidth]{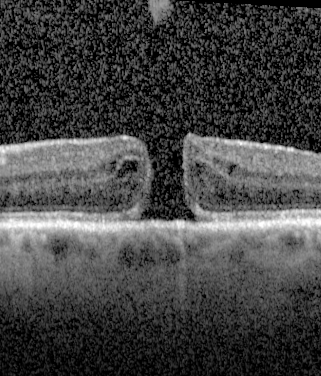}
    & \includegraphics[trim=50 110 50 100, clip, width=.166\linewidth]{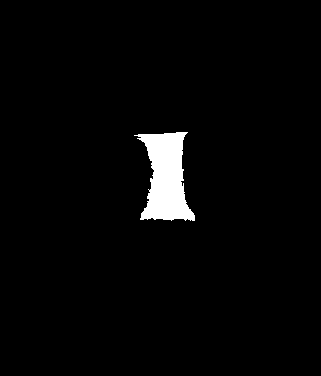}
    & \includegraphics[trim=50 110 50 100, clip, width=.166\linewidth]{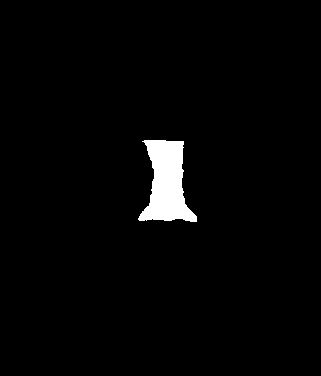}
    & \includegraphics[trim=50 110 50 100, clip, width=.166\linewidth]{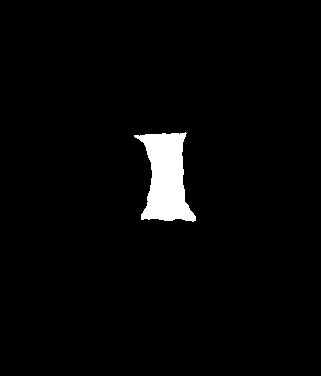}
    & \includegraphics[trim=50 110 50 100, clip, width=.166\linewidth]{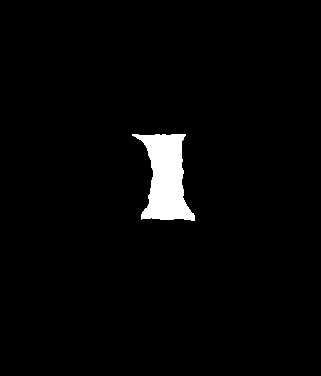}
    & \includegraphics[trim=50 110 50 100, clip, width=.166\linewidth]{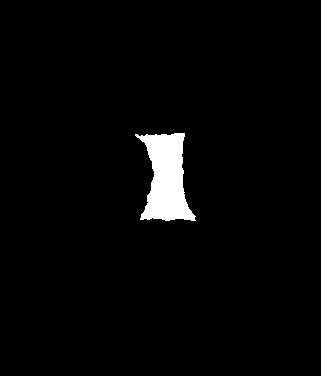} \\
    \includegraphics[trim=50 150 50 50, clip, width=.166\linewidth]{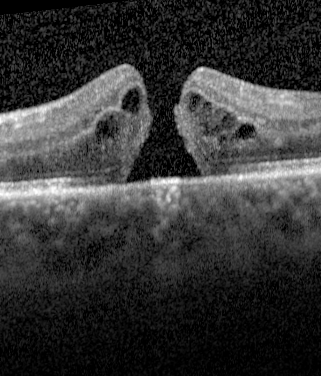}
    & \includegraphics[trim=50 150 50 50, clip, width=.166\linewidth]{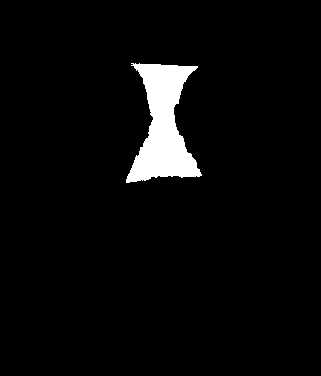}
    & \includegraphics[trim=50 150 50 50, clip, width=.166\linewidth]{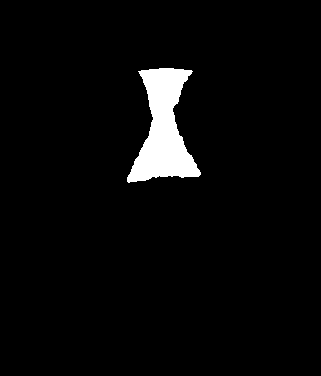}
    & \includegraphics[trim=50 150 50 50, clip, width=.166\linewidth]{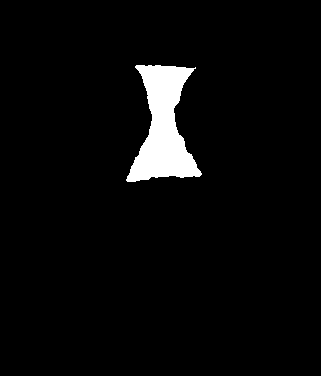}
    & \includegraphics[trim=50 150 50 50, clip, width=.166\linewidth]{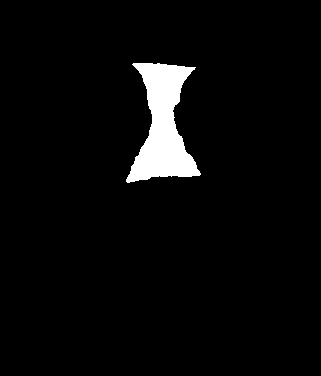}
    & \includegraphics[trim=50 150 50 50, clip, width=.166\linewidth]{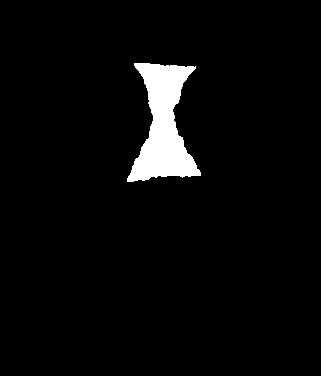} \\
    \includegraphics[trim=50 140 50 75, clip, width=.166\linewidth]{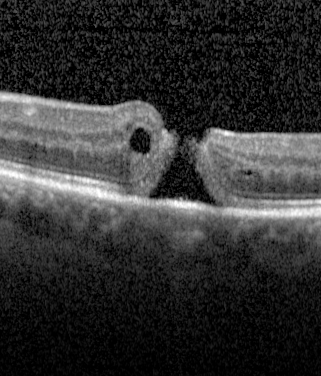}
    & \includegraphics[trim=50 140 50 75, clip, width=.166\linewidth]{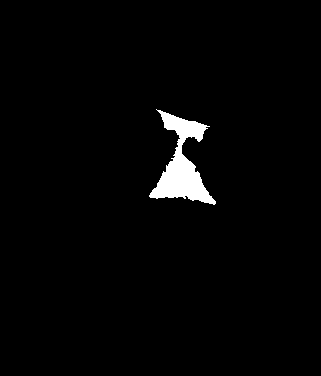}
    & \includegraphics[trim=50 140 50 75, clip, width=.166\linewidth]{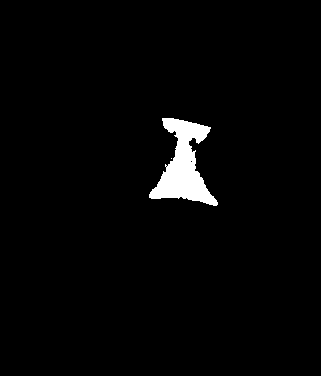}
    & \includegraphics[trim=50 140 50 75, clip, width=.166\linewidth]{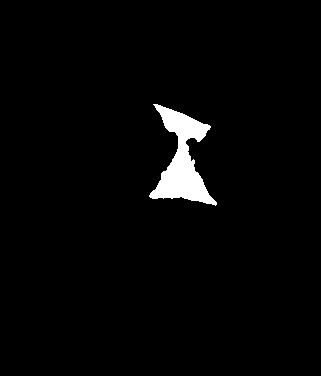}
    & \includegraphics[trim=50 140 50 75, clip, width=.166\linewidth]{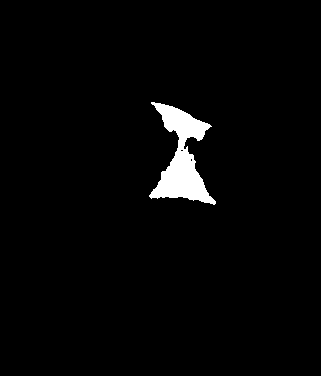}
    & \includegraphics[trim=50 140 50 75, clip, width=.166\linewidth]{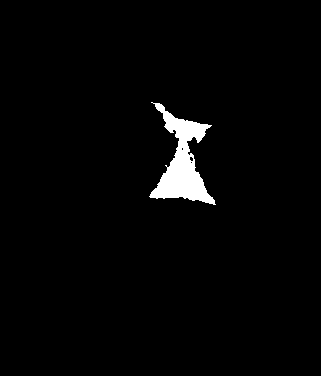} \\
    OCT scan & Ground truth & Nasrulloh & $M_{1}$ & $M_{2}$ & $M_{3}$ \\
  \end{tabular}
  \caption{Qualitative sample output of our trained macular hole model ($M_{1}$) compared with the ground truth, the state-of-the-art automated approach (Nasrulloh)~\cite{Nasrulloh2018}, the residual model ($M_{2}$) and DeepMind's model ($M_{3}$). For clarity, we zoomed in on the predicted regions.}
  \label{fig:maculargood}
\end{figure*}

\begin{table}[h!]
  \setlength\tabcolsep{.006\linewidth}
  \centering
  \caption{3D and 2D segmentation output of model $M_{1}$ (Proposal) on unseen test images along with ground truth}
  \begin{tabular}{ccc}\toprule
3D segmentation & 2D segmentation & 2D ground truth \\ \midrule
                           
$\begin{array}{l}\includegraphics[width=.34\linewidth]{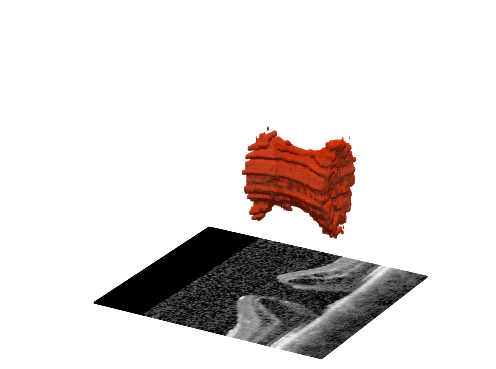}\end{array}$&$\begin{array}{l}\includegraphics[width=.23\linewidth]{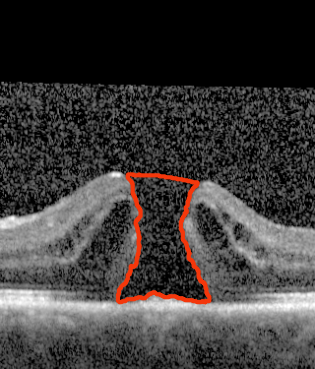}\end{array}$&$\begin{array}{l}
\includegraphics[width=.23\linewidth]{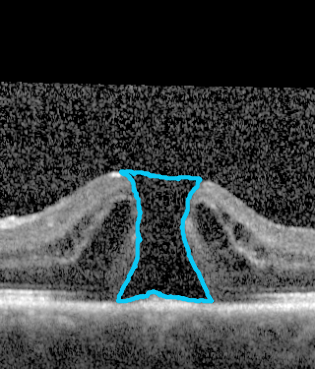}\end{array}$ \\                                                                                                                                                

$\begin{array}{l}\includegraphics[width=.34\linewidth]{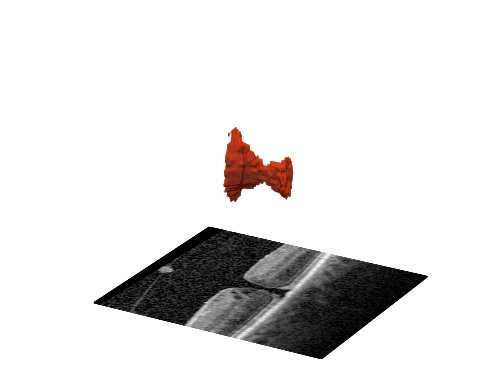}\end{array}$&$\begin{array}{l}\includegraphics[width=.23\linewidth]{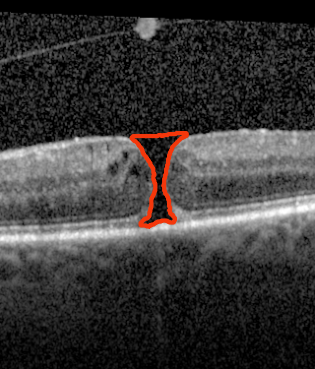}\end{array}$&$\begin{array}{l}
\includegraphics[width=.23\linewidth]{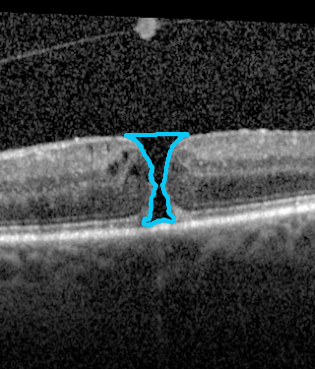}\end{array}$ \\

$\begin{array}{l}\includegraphics[width=.34\linewidth]{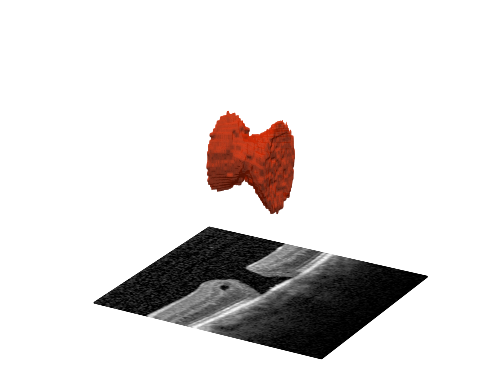}\end{array}$&$\begin{array}{l}\includegraphics[width=.23\linewidth]{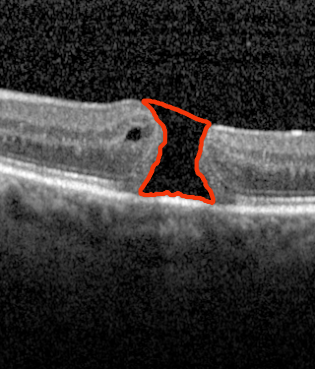}\end{array}$&$\begin{array}{l}
\includegraphics[width=.23\linewidth]{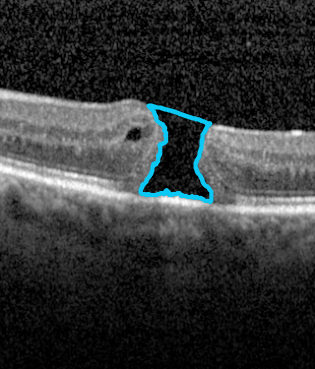}\end{array}$ \\                                                                                                                                                
                                                                                                                                                                                                                                              
    \bottomrule                                                                                                                                                                                                                               
  \end{tabular}
  \label{fig:mh3d}         
\end{table}                                     

\section{Materials}
OCT images were exported from a Heidelberg SPECTRALIS (Heidelberg, Germany) machine.
A high density central horizontal scanning protocol with 29-30 micron line spacing was used in the central 15 by 5 degrees.
The individual OCT line scans had spacing which varied slightly between images but it was typically 5.47 microns per pixel on the $x$-axis and 3.87 microns per pixel on the $y$-axis.
All scans used a 16 automatic real-time setting enabling multisampling and noise reduction over 16 images. 

All images were cropped to the same size and unnecessary information such as the fundus image were removed.
Annotations were created by a mixture of clinicians and image experts using a 3D image annotation tool.
Pixels on each slice of the OCT scan which represented macular hole were highlighted.
There were 85 (image, annotation) pairs in the training dataset, 56 after combining annotations from multiple authors.
There were 22 pairs in the validation dataset and 9 in the unseen test set.

Originally we had three annotations for each OCT image in the unseen test set.
However, due to inconsistencies between authors, we combined all ground truths into a single ground truth per image.
To do this, we used a voting system, where if $\frac{2}{3}$ of the authors had annotated a voxel, that voxel was annotated in the resultant ground truth.
All images and ground truths at full size had dimensions $321 \times 376 \times 49$.

\section{Results}
In this section, we evaluate our results both qualitatively and quantitatively.

\subsection{Qualitative Results}

The qualitative results of running the trained macular hole models are generally quite close to the ground truth, as seen in \figurename~\ref{fig:maculargood}.
In general, predictions from the models are closer to the ground truth than the state-of-the-art automated approach.
3D visualizations of the output of our proposed model can be seen in \tablename~\ref{fig:mh3d}.

\subsection{Quantitative Results}
\label{sec:quant}

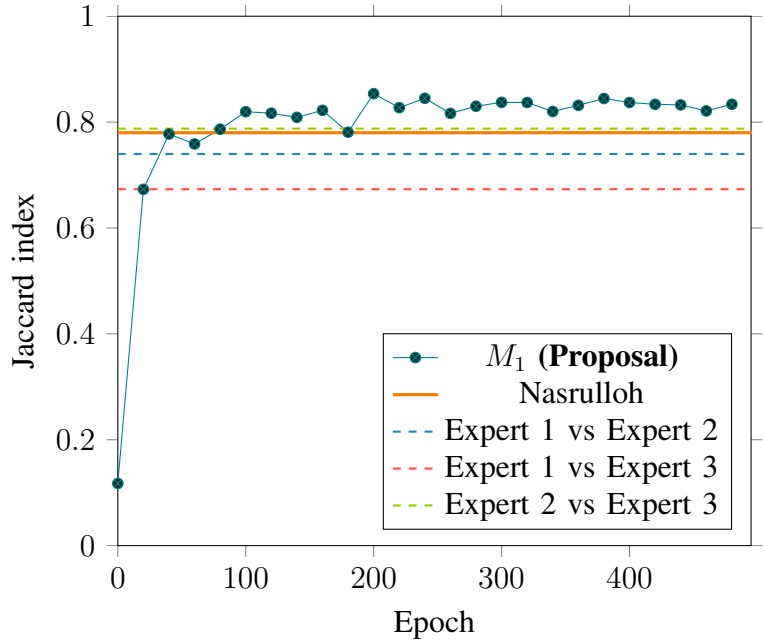
\begin{figure}[ht]
    \centering

    \begin{tikzpicture}[thick,scale=1.0, every node/.style={scale=0.8}]
    \tikzstyle{every node}=[font=\normalsize]
    \begin{axis}[
        xlabel={Epoch},
        ylabel={Jaccard index},
        xmax=495,
        ymin=0.0,
        ymax=1.0,
        enlarge x limits=false,
        cycle list name=my exotic colorlist,
        legend pos=south east
        ]

    \addplot+[] table [
        x=epoch,
        y=jaccard,
        col sep=comma
        ] {results/500/U-Net_3D.csv};
    \addlegendentry{$M_{1}$ \textbf{(Proposal)}}

    \addplot[mark=none, orange, very thick] coordinates {(0,0.78) (500,0.78)};
    \addlegendentry{Nasrulloh}

    \addplot[mark=none, cyan!60!black, dashed, thick] coordinates {(0,0.7397858464940825) (500,0.7397858464940825)};
    \addlegendentry{Expert 1 vs Expert 2}

    \addplot[mark=none, red!70!white, dashed, thick] coordinates {(0,0.6732643550405537) (500,0.6732643550405537)};
    \addlegendentry{Expert 1 vs Expert 3}

    \addplot[mark=none, lime!80!black, dashed, thick] coordinates {(0,0.7877348922575328) (500,0.7877348922575328)};
    \addlegendentry{Expert 2 vs Expert 3}

    \end{axis}
    \end{tikzpicture}

    \caption{Average Jaccard index of our proposed model ($M_{1}$) over 3 runs on the unseen test set as the model was trained (higher is better). We see that the model achieves significantly better results than the state-of-the-art automated approach (Nasrulloh) and expert agreement.}
    \label{fig:jacovertime}
\end{figure}

\begin{table}[ht]
    \centering
    \setlength\tabcolsep{3.5pt} 
    \caption{Jaccard index comparison between the state-of-the-art~\cite{Nasrulloh2018} automated approach and tested models (mean and standard deviation over three runs except for state-of-the-art which is deterministic)}
\begin{tabular}{@{}lccccc@{}} \toprule
Image&Nasrulloh&$M_{1}$ (Proposal)&$M_{2}$ (Residual)&$M_{3}$ (DeepMind)\\ \midrule

Image 1&0.714&$0.865 \pm 0.009$&$0.868 \pm 0.002$&$0.832 \pm 0.006$&\\
Image 2&0.743&$0.891 \pm 0.02$&$0.887 \pm 0.014$&$0.893 \pm 0.012$&\\
Image 3&0.772&$0.887 \pm 0.004$&$0.885 \pm 0.002$&$0.872 \pm 0.006$&\\
Image 4&0.811&$0.895 \pm 0.012$&$0.884 \pm 0.001$&$0.894 \pm 0.006$&\\
Image 5&0.787&$0.894 \pm 0.005$&$0.901 \pm 0.003$&$0.875 \pm 0.014$&\\
Image 6&0.678&$0.804 \pm 0.008$&$0.815 \pm 0.007$&$0.765 \pm 0.006$&\\
Image 7&0.845&$0.907 \pm 0.002$&$0.905 \pm 0.004$&$0.893 \pm 0.009$&\\
Image 8&0.874&$0.874 \pm 0.012$&$0.862 \pm 0.002$&$0.893 \pm 0.006$&\\
Image 9&0.787&$0.869 \pm 0.019$&$0.853 \pm 0.008$&$0.835 \pm 0.007$&\\
\midrule
Mean&$0.779$&$0.876 \pm 0.012$&$0.874 \pm 0.006$&$0.861 \pm 0.008$&\\
  \bottomrule
\end{tabular}
    \label{comparison}
\end{table}

\begin{table}[ht]
    \centering
    \setlength\tabcolsep{3.5pt} 
    \caption{Detailed statistics of model $M_{1}$ (Proposal) (DSC refers to dice similarity coefficient, AVD refers to absolute volume difference and AP refers to average precision)}
    \begin{tabular}{@{}lccccc@{} } \toprule
Image&Precision&Recall&DSC&AVD&AP\\ \midrule
Image 1&$0.93 \pm 0.009$&$0.926 \pm 0.012$&$0.928 \pm 0.005$&$1352 \pm 908.357$&$0.862 \pm 0.01$\\
Image 2&$0.954 \pm 0.008$&$0.931 \pm 0.014$&$0.942 \pm 0.011$&$1379 \pm 369.396$&$0.889 \pm 0.021$\\
Image 3&$0.949 \pm 0.003$&$0.931 \pm 0.003$&$0.94 \pm 0.002$&$2308 \pm 517.533$&$0.885 \pm 0.004$\\
Image 4&$0.974 \pm 0.005$&$0.917 \pm 0.016$&$0.945 \pm 0.007$&$5293 \pm 1817.849$&$0.895 \pm 0.011$\\
Image 5&$0.915 \pm 0.012$&$0.974 \pm 0.007$&$0.944 \pm 0.003$&$2320 \pm 763.08$&$0.892 \pm 0.005$\\
Image 6&$0.848 \pm 0.003$&$0.94 \pm 0.007$&$0.891 \pm 0.005$&$1911 \pm 80.168$&$0.797 \pm 0.009$\\
Image 7&$0.965 \pm 0.002$&$0.938 \pm 0.001$&$0.951 \pm 0.001$&$1564 \pm 157.11$&$0.906 \pm 0.002$\\
Image 8&$0.898 \pm 0.006$&$0.971 \pm 0.008$&$0.933 \pm 0.007$&$4544 \pm 155.656$&$0.872 \pm 0.013$\\
Image 9&$0.917 \pm 0.016$&$0.943 \pm 0.012$&$0.93 \pm 0.011$&$1186 \pm 906.832$&$0.865 \pm 0.02$\\
    \bottomrule
    \end{tabular}
    \label{detailed}
\end{table}
\figurename~\ref{fig:jacovertime} shows how the Jaccard index improves as $M_{1}$ is trained and we can see that after 200 epochs it surpasses the performance of the state-of-the-art automated approach and expert agreement.
All of the trained macular hole models perform very well compared to the state-of-the-art automated approach~\cite{Nasrulloh2018} as we see in \tablename~\ref{comparison}. 
Despite $M_{1}$ having by far the fewest parameters, it achieves performance which is similar to the higher-capacity models, and in some cases surpasses them.
Further results in \tablename~\ref{detailed} show that $M_{1}$ performs consistently well under other standard segmentation quality measures such as dice similarity coefficient.

\section{Conclusion}
All of the models tested exceeded the performance of the state-of-the-art automated approach which is a level set method.
It is clear that deep learning methods allow for the generation of segmentations which are closer to what humans provide.
Despite $M_{3}$ having 90 times the parameters of $M_{1}$, $M_{1}$ gives excellent qualitative and quantitative results which are of a similar quality to $M_{3}$.
$M_{1}$'s performance exceeded expert agreement by a Jaccard index of $0.13-0.20$.
As $M_{1}$ is the smallest model, it requires the least amount of resources to run.
$M_{1}$ is also a quick model to run, requiring only one pass through the whole 3D image, whereas $M_{3}$ requires one pass per slice.
Once trained, $M_{1}$ is capable of segmenting an OCT image in less than one second.
In contrast, the state-of-the-art automated method requires minutes to run~\cite{Nasrulloh2018}.
For these reasons, $M_{1}$ is the best candidate to form the basis of future studies in a clinical setting.
These findings show that careful tuning and in some cases architectural simplification can, for some simple task distributions, be as effective as very deep residual designs. \\

\section*{Conflict of Interest and Acknowledgements}
In accordance with his ethical obligation as a researcher, Jonathan Frawley reports that he received funding for his PhD from Intogral Ltd.
Some of the work described was developed as part of his work as an employee at Intogral Ltd.
Intogral Ltd also provided annotations created by non-clinicians.
Data and annotations by the clinician for this project were kindly provided by Maged Habib, Caspar Geenen and David H.~Steel of the Sunderland Eye Infirmary, South Tyneside and Sunderland NHS Foundation Trust, UK.
All images were collected as part of routine care and anonymised.

\bibliographystyle{IEEEtran}
\bibliography{main}

\begin{thebibliography}{10}
\providecommand{\url}[1]{#1}
\csname url@samestyle\endcsname
\providecommand{\newblock}{\relax}
\providecommand{\bibinfo}[2]{#2}
\providecommand{\BIBentrySTDinterwordspacing}{\spaceskip=0pt\relax}
\providecommand{\BIBentryALTinterwordstretchfactor}{4}
\providecommand{\BIBentryALTinterwordspacing}{\spaceskip=\fontdimen2\font plus
\BIBentryALTinterwordstretchfactor\fontdimen3\font minus
  \fontdimen4\font\relax}
\providecommand{\BIBforeignlanguage}[2]{{%
\expandafter\ifx\csname l@#1\endcsname\relax
\typeout{** WARNING: IEEEtran.bst: No hyphenation pattern has been}%
\typeout{** loaded for the language `#1'. Using the pattern for}%
\typeout{** the default language instead.}%
\else
\language=\csname l@#1\endcsname
\fi
#2}}
\providecommand{\BIBdecl}{\relax}
\BIBdecl

\bibitem{ali2017incidence}
F.~S. Ali, J.~D. Stein, T.~S. Blachley, S.~Ackley, and J.~M. Stewart,
  ``Incidence of and risk factors for developing idiopathic macular hole among
  a diverse group of patients throughout the {U}nited {S}tates,'' \emph{JAMA
  Ophthalmology}, vol. 135, no.~4, pp. 299--305, 2017.

\bibitem{mccannel2009population}
C.~A. McCannel, J.~L. Ensminger, N.~N. Diehl, and D.~N. Hodge,
  ``Population-based incidence of macular holes,'' \emph{Ophthalmology}, vol.
  116, no.~7, pp. 1366--1369, 2009.

\bibitem{hee1995optical}
M.~R. Hee, C.~A. Puliafito, C.~Wong, J.~S. Duker, E.~Reichel, J.~S. Schuman,
  E.~A. Swanson, and J.~G. Fujimoto, ``Optical coherence tomography of macular
  holes,'' \emph{Ophthalmology}, vol. 102, no.~5, pp. 748--756, 1995.

\bibitem{goldberg2014optical}
R.~A. Goldberg, N.~K. Waheed, and J.~S. Duker, ``Optical coherence tomography
  in the preoperative and postoperative management of macular hole and
  epiretinal membrane,'' \emph{British Journal of Ophthalmology}, vol.~98, no.
  Suppl 2, pp. ii20--ii23, 2014.

\bibitem{madi2016optimal}
H.~A. Madi, I.~Masri, and D.~H. Steel, ``Optimal management of idiopathic
  macular holes,'' \emph{Clinical Ophthalmology (Auckland, NZ)}, vol.~10,
  p.~97, 2016.

\bibitem{chen2020macular}
Y.~Chen, A.~V. Nasrulloh, I.~Wilson, C.~Geenen, M.~Habib, B.~Obara, and D.~H.
  Steel, ``Macular hole morphology and measurement using an automated
  three-dimensional image segmentation algorithm,'' \emph{BMJ Open
  Ophthalmology}, vol.~5, no.~1, p. e000404, 2020.

\bibitem{murphy2020}
\BIBentryALTinterwordspacing
D.~C. Murphy, A.~V. Nasrulloh, C.~Lendrem, S.~Graziado, M.~Alberti, M.~{la
  Cour}, B.~Obara, and D.~H. Steel, ``Predicting postoperative vision for
  macular hole with automated image analysis,'' \emph{Ophthalmology Retina},
  2020. [Online]. Available:
  \url{http://www.sciencedirect.com/science/article/pii/S2468653020302311}
\BIBentrySTDinterwordspacing

\bibitem{steel2020factors}
D.~H. Steel, P.~H. Donachie, G.~W. Aylward, D.~A. Laidlaw, T.~H. Williamson,
  and D.~Yorston, ``Factors affecting anatomical and visual outcome after
  macular hole surgery: findings from a large prospective {UK} cohort,''
  \emph{Eye}, pp. 1--10, 2020.

\bibitem{Nasrulloh2018}
A.~Nasrulloh, C.~Willcocks, P.~T. Jackson, C.~Geenen, M.~S. Habib, D.~H. Steel,
  and B.~Obara, ``Multi-scale segmentation and surface fitting for measuring
  3{D} macular holes,'' \emph{IEEE Transactions on Medical Imaging}, vol.~37,
  no.~2, pp. 580--589, 2018.

\bibitem{Lindsay_2020}
\BIBentryALTinterwordspacing
G.~W. Lindsay, ``Convolutional neural networks as a model of the visual system:
  Past, present, and future,'' \emph{Journal of Cognitive Neuroscience}, p.
  1–15, Feb 2020. [Online]. Available:
  \url{http://dx.doi.org/10.1162/jocn_a_01544}
\BIBentrySTDinterwordspacing

\bibitem{ronneberger2015u}
O.~Ronneberger, P.~Fischer, and T.~Brox, ``U-{N}et: convolutional networks for
  biomedical image segmentation,'' in \emph{Medical Image Computing and
  Computer-Assisted Intervention}, vol. 9351, 2015, pp. 234--241.

\bibitem{frawleybibe2020}
J.~Frawley, C.~Willcocks, H.~Maged, G.~Caspar, D.~H. Steel, and B.~Obara,
  ``Segmentation of macular edema datasets with small residual 3d u-net
  architectures.'' in \emph{20th IEEE International Conference on
  BioInformatics and BioEngineering}.\hskip 1em plus 0.5em minus 0.4em\relax
  IEEE, 2020, pp. 582--587.

\bibitem{cciccek20163d}
{\"O}.~{\c{C}}i{\c{c}}ek, A.~Abdulkadir, S.~S. Lienkamp, T.~Brox, and
  O.~Ronneberger, ``3{D} {U}-{N}et: learning dense volumetric segmentation from
  sparse annotation,'' in \emph{{I}nternational {C}onference on {M}edical
  {I}mage {C}omputing and {C}omputer-{A}ssisted {I}ntervention}, 2016, pp.
  424--432.

\bibitem{he2016identity}
K.~He, X.~Zhang, S.~Ren, and J.~Sun, ``Identity mappings in deep residual
  networks,'' in \emph{European Conference On Computer Vision}, 2016, pp.
  630--645.

\bibitem{de2018clinically}
J.~De~Fauw, J.~R. Ledsam, B.~Romera-Paredes, S.~Nikolov, N.~Tomasev,
  S.~Blackwell \emph{et~al.}, ``Clinically applicable deep learning for
  diagnosis and referral in retinal disease,'' \emph{Nature Medicine}, vol.~24,
  no.~9, p. 1342, 2018.

\bibitem{taha2015metrics}
A.~A. Taha and A.~Hanbury, ``Metrics for evaluating 3d medical image
  segmentation: analysis, selection, and tool,'' \emph{BMC medical imaging},
  vol.~15, no.~1, pp. 1--28, 2015.

\bibitem{pytorch}
A.~Paszke, S.~Gross, F.~Massa \emph{et~al.}, ``Py{T}orch: an imperative style,
  high-performance deep learning library,'' in \emph{Advances in Neural
  Information Processing Systems 32}, 2019, pp. 8024--8035.

\bibitem{kingma2014adam}
D.~P. Kingma and J.~Ba, ``Adam: {A} method for stochastic optimization,'' in
  \emph{International Conference on Learning Representations}, Y.~Bengio and
  Y.~LeCun, Eds., 2015.

\end{thebibliography}

\end{document}